\documentclass[12pt]{article}
\pdfoutput=1

\usepackage{graphicx}
\usepackage{epstopdf}
\usepackage[body={17.5cm, 21cm},right=2cm]{geometry}
\usepackage{amssymb}
\usepackage{amsmath}
\usepackage{xcolor,graphicx}

\newcommand\eea{\end{eqnarray}}
\newcommand\bea{\begin{eqnarray}}

\newcommand\mpl{M_{P}}

\def\beq{\begin{equation}}
\def\eeq{\end{equation}}

\def\e{{\rm e}}

\newcommand{\be}{\begin{equation}}
\newcommand{\ee}{\end{equation}}
\newcommand{\ba}{\begin{align}}
\newcommand{\ea}{\end{align}}
\newcommand{\bg}{\begin{gather}}
\newcommand{\eg}{\end{gather}}
\newcommand{\bseq}{\begin{subequations}}
\newcommand{\eseq}{\end{subequations}}

\renewcommand{\ln}{\mathop{\rm ln}\nolimits}

\newcommand{\bx}{\bf x}
\newcommand{\bk}{\bf k}

\begin{document}

\begin{flushright}
SU-ITP-11/43 ~~~ SLAC-PUB-14551
\end{flushright}

\vspace{3mm}
\vspace{0.5cm}
\begin{center}

\def\thefootnote{\fnsymbol{footnote}}

{\Large \bf New Sources of Gravitational Waves during Inflation}
\\[0.5cm]
{\large  Leonardo Senatore$^1$, Eva Silverstein$^1$, and Matias Zaldarriaga$^2$}
\\[0.5cm]

{\small \textit{$^{1}$
Department of Physics and SLAC \\
Stanford University, Stanford, CA 94305, USA}}

\vspace{.2cm}

{\small \textit{$^{2}$ Institute for Advanced Study\\Einstein Drive\\ Princeton, NJ 08540}}

\end{center}

\vspace{.8cm}

\hrule \vspace{0.3cm}
{\small  \noindent \textbf{Abstract} \\[0.3cm]
\noindent
We point out that detectable inflationary tensor modes can be generated by particle or string sources produced during inflation, consistently with the requirements for inflation and constraints from scalar fluctuations.  We show via examples that this effect can dominate over the contribution from quantum fluctuations of the metric, occurring even when the inflationary potential energy is too low to produce a comparable signal.  Thus a detection of tensor modes from inflation does not automatically constitute a determination of the inflationary Hubble scale.

\vspace{0.5cm}  \hrule
\def\thefootnote{\arabic{footnote}}
\setcounter{footnote}{0}




\vspace{.8cm}

\tableofcontents

\vspace{.8cm}

\section{Introduction and Motivations}

Gravitational radiation, detectable through B mode polarization in the CMB \cite{Bmodes}, provides an important handle on primordial cosmology.  In the context of inflation, quantum fluctuations of the tensor modes in the metric yield a power spectrum of the form
\be\label{inflGW}
 \langle h^s_{\bf k}h^{s'}_{\bf k'} \rangle = (2\pi)^3 \delta({\bf k}+{\bf k'})\delta^{ss'} {\cal P}_h \ , \qquad P_h = \frac{1}{k^3}\frac{H^2}{M_P^2}\ ,
\ee
whose amplitude is given directly by the scale of the inflaton potential $V(\phi)\sim H^2 M_P^2$ (the index $s$ here labels the polarization state) .
For this reason, a detection of primordial B modes is often identified with a measurement of the Hubble scale $H$ during inflation. It is also related to the range of the inflaton \cite{Lyth}\ in Planck units,  providing an ultraviolet-sensitive observable.  Observational projects sensitive to B modes are expected to test GUT-scale inflation and Planck-scale field ranges in the relatively near term.
To be detectable at least in near-term observational projects,\footnote{See \cite{Farhang:2011ud}\ and references therein for a recent treatment of observational expectations.} the amplitude must satisfy
\beq\label{detecth}
h\sim h_k k^{3/2} \gtrsim 10^{-6}
\eeq

In this paper we will consider additional sources of gravitational waves which may be present during inflation.\footnote{See \cite{Chialva:2010jt}\ for an interesting analysis of gravitational waves from phase transitions during inflation, and \cite{Chungetal}\ for earlier work on some effects of particle production during inflation, in the regime of parametric resonance \cite{KLS}.}
The inflaton $\phi$ might generically be expected to couple to other degrees of freedom $X$ and produce excited states of these sectors as it rolls through points in field space where they become light.
These excited degrees of freedom can source gravitational waves (GWs), which freeze out as they cross the horizon.  A basic question is to what extent these additional sources can produce detectable primordial gravity waves, and how their amplitude $h_X$ and other properties compare to those produced by the basic process (\ref{inflGW}).

Of course the energy density $\rho_X$ in the $X$ sector dilutes away during inflation because of the exponential expansion.  A single production event would lead to a scale-dependent feature in the GW spectrum, an interesting possibility in itself.
However, during inflation $\phi$ may well encounter multiple points with new light degrees of freedom.  For example, in the mechanism \cite{monodromy,trapped}\ for inflation along angular (axion) directions which are typically extended via monodromy, this is automatic: any such production process is repeated at short intervals because of an underlying circle in field space.  This repetition of the process replenishes the supply of extra modes as inflation dilutes them.  The requirement of reheating involves coupling of the inflaton to other sectors after exit, and in \cite{monodromy, trapped}\ it would be interesting to study the implications of this for earlier particle production events.  In other cases with a sufficiently rich spectrum of fields, repeated production events may also occur, and in general it is interesting to consider possible observational consequences.

With this motivation, in this note we estimate $h_X$ for particles and for strings produced when their mass or tension depends sufficiently strongly on the inflaton.  We find that the amplitude of the GWs generated by various processes including particle/string production, Bremsstrahlung, string oscillations, and decays can compete (and in some cases exceed) the contribution (\ref{inflGW}). This is of interest for two reasons:

$\bullet$ It indicates that even within the context of inflation, the observation of a scale invariant spectrum of B modes does not automatically constitute a direct measurement of the inflationary potential.  This is relevant for the goal of having a systematic treatment of inflationary mechanisms and signatures.

$\bullet$ It provides a new regime in which to look for observational signatures of exotic sources.\footnote{The interesting possibility of detecting {\it post-inflationary} exotics such as cosmic strings has been studied extensively; here we are concerned with a distinct window of potential signatures.}

$\bullet$  It introduces an additional (model-dependent) signature of some classes of inflationary mechanisms such as \cite{monodromy}.\footnote{This mechanism makes predictions for the amplitude of GWs from tensor fluctuations and for the tilt of the spectrum, and depending on the model parameters can lead to additional signatures which are more detailed, such as oscillations in the power spectrum and resonant non-Gaussianity \cite{monodromy,otherOsc}.  The present work introduces another model-dependent signature.}

The paper is organized as follows.  We start in the next section by deriving the basic requirements for a detectable signal and how this fits well within the basic bounds on energy densities of extra sectors during inflation.
Then we review and apply the standard calculation of gravitational waves produced by stress energy sources, giving several examples of particle and string sources which generate a competitive signal.
In an appendix we will review particle production in the presence of a time-dependent mass, and describe the salient features we will need of string production in the presence of a time-dependent tension.

\section{Basic checks}
\label{sec:basicchecks}

A basic requirement of the additional sources is that their energy density $\rho_X$ be subdominant to the inflationary potential energy:
\beq\label{energy}
\rho_X\ll V\sim H^2 M_P^2
\eeq
by at least a factor of the slow roll parameter $\epsilon\sim \frac{\dot H^2}{H^4}\lesssim 10^{-1}$.  Given that they are produced by the rolling scalar field, the energy density $\rho_X$ in the additional sources will be at most of the order
\beq
\dot\phi^2\sim \epsilon H^2 M_P^2.
\eeq
It will be of this order if the inflaton dumps a significant fraction of its kinetic energy into the $X$ sector, as can happen naturally on a steep potential \cite{trapped}.  The fraction $f\equiv \rho_X/H^2M_P^2\le\epsilon$ of the total energy density which is carried by the sources will figure into our estimates below for the strength of the GW signal.

Once produced by the extra sources $X$, the additional gravitational waves satisfy the standard equation of motion in the inflationary near-de Sitter background, forming linear combinations
\beq\label{hankel}
h_k(t) \approx
 A_{1k} \left(i+\frac{k}{a(t)H}\right)e^{ik/a(t)H}+ A_{2k} \left(-i+\frac{k}{a(t)H}\right)e^{-ik/a(t)H} \ ,
\eeq
with $A_{(1,2)k}$ coefficients determined by the initial conditions, which freeze out when $k/a(t)$ reaches the Hubble scale $H$ as usual.  (This solution ignores the slow time-variation of $H$.)  Gravitational waves that are produced inside the Hubble patch decrease in amplitude by a factor of $H/(k/a_i)$ before freezing out, where $a_i$ is the scale factor at the initial time of production of the mode and $k/a_i\sim\omega_i$ its physical momentum at that time.

Gravitational waves of initial frequency $\omega_i$
make up an energy density at freeze out of order
\beq\label{energetics}
\rho_{GW}\sim \dot h(t)^2 M_P^2|_{freezeout} \sim \omega_i^2 M_P^2h_i^2 \left(\frac{H}{\omega_i}\right)^4
\sim \rho_i \left(\frac{H}{\omega_i}\right)^4
\eeq
where $\rho_i$ is the initial energy density contained in the GWs and $h_i$ their amplitude at the time of their production; in the last factor we took into account the redshifting noted above of the modes and of the frequencies.  We are interested in whether the resulting frozen out modes can be competitive with those from GUT-scale inflationary theory and visible in near-term observations.  Comparison with (\ref{detecth}) shows that for detectability
and consistency with the $\epsilon$ condition (\ref{energy}) we must require
\beq\label{amplitudecondition}
10^{-6}\le h_i \frac{H}{\omega_i}\le \sqrt{\epsilon}\left(\frac{H}{\omega_i}\right)^2 ~~ \Rightarrow ~~ \omega_i\le 10^3 \epsilon^{1/4} H
\eeq
Thus the possibility of observable GWs sourced by the $X$ sector is not immediately excluded by any simple consideration of energetics.
In order to determine if this possibility is viable, we must work out the spectrum of frequencies $\omega_i$ in concrete examples.  In particular, as we will discuss further below, low-frequency GWs can be suppressed by interference from multiple scattering events in a dense gas of $X$ particles or strings.
It is interesting to note that if the $X$ sector degrees of freedom decayed into high-frequency GWs with a typical frequency $\omega_i$ of order $\sqrt{\dot\phi_0^2}=\sqrt{\epsilon}HM_P$, the condition (\ref{amplitudecondition}) translates into the condition $\frac{H}{M_P}\ge 10^{-6}$, which is to say that the effect would still be marginally competitive.

We should also emphasize that the $X$ sector can in general also emit scalar perturbations $\delta\phi$; since its production arises from its coupling to the rolling inflaton field $\phi$, at the time it is created there is a nontrivial coupling between $X$ and $\delta \phi$.  Its coupling to $\delta\phi$ at later times is model dependent (related to the functional form of the $\phi$-dependent mass or tension).  In each of the examples below, we will determine the strength of the scalar perturbations and estimate their non-Gaussianity to ensure that they are consistent with phenomenological constraints.

It will be useful to rephrase the condition for detectability in terms of the energy density contained in gravitational waves.  In general, from (\ref{energetics}), this requires
\beq\label{detect}
h^2|_{freezeout}\sim \frac{\rho_{GW}}{H^2M_P^2} \ge 10^{-12}
\eeq
and below we will estimate this quantity for particle and string sectors $X$.

\section{Gravitational Wave Sources}

In this section, we will start by reviewing the standard derivation of gravitational radiation, following the comprehensive treatment in \cite{Weinberg}, which includes a detailed analysis of Bremsstrahlung.  We will wish to generalize this analysis in several ways.  In particular, we will consider decays and production events as well as Bremsstrahlung.  Also, we would like to include the effects of the inflaton field $\phi$ coupling to particle or string sources, with $\phi$ determining their mass or tension.

\subsection{General setup}
\label{subsec:generalsetup}

Let us begin by briefly collecting some of the basic results on gravitational radiation, and set up our system.  We will shortly make simple estimates in special cases, but it is worthwhile to first lay out the general problem.
Given the stress-energy of sources, at the linearized level one obtains tensor perturbations \cite{Weinberg}
\beq
h_{\mu\nu}=\frac{4}{8\pi M_P^2}\int \frac{d^3\bx'}{|\bx-\bx'|}S_{\mu\nu}(\bx',\omega)e^{-i\omega t+i\omega |\bx-\bx'|}+c.c.  \eeq
where $S_{\mu\nu}=T_{\mu\nu}-\frac{1}{2}\eta_{\mu\nu}T^\lambda_\lambda(\bx,\omega)$.  It is convenient to work with the Fourier transform of the stress-energy tensor, and one finds a result for the total energy emitted per solid angle
\beq
\frac{dE}{d\Omega}=\frac{2}{8\pi M_P^2}\int_0^\infty d\omega \omega^2\left(T^{\lambda\nu *}(\bk,\omega)T_{\lambda\nu}(\bk,\omega)-\frac{1}{2}|T^\lambda_\lambda(\bk,\omega)|^2\right)\ .
\eeq
A particularly simple situation to consider is one in which particles of fixed (time-independent) masses scatter and emit GWs through Bremsstrahlung.  As derived in \cite{Weinberg}, that leads to a GW signature with total energy
\beq\label{BremGW}
\left(\frac{dE}{d\Omega d\omega}\right)=\frac{\omega^2}{2\pi^2M_P^2}\sum_{N,M}\frac{\eta_N\eta_M}{(P_N\cdot k)(P_M\cdot k)}\left[(P_N\cdot P_M)^2-\frac{1}{2}m_N^2m_M^2\right]
\eeq
where we have taken the limit that the wavelength of the emitted Bremsstrahlung radiation is long compared to the scattering time and the time between scattering events.
Here $N,M$ index particles with momentum $P_N,P_M$, and $\eta_N,\eta_M$ are $\pm 1$ depending on whether the particle is ingoing or outgoing in a given event.  In particular, if there is no scattering, so that the incoming and out going momenta are the same, (\ref{BremGW}) gives zero via the cancelations arising from the $\eta_N,\eta_M$ factors.

For processes occurring well within the Hubble scale $H^{-1}$, stress-energy is conserved to a good approximation, shared between particle or string sources and the inflaton field.  Before considering specific examples, let us briefly set up the full problem.  The classical action is
\bea\label{actionfull}
{\cal S} &=&\int d^4 x\sqrt{-g}\left(\frac{1}{2} M_P^2 {\cal R} +{\cal L}_\phi \right) + {\cal S}_X + {\cal S}_{XY} + {\cal S}_{Y}\\
{\cal S}_X &=& -\sum_{p}\int d^4 x\int d\tau \delta^{(4)}(x^\mu-x_p^\mu(\tau)) m(\phi(t,\bx))\sqrt{-g_{\mu\nu}(x_p(\tau))\frac{dx^\mu(\tau)}{d\tau} \frac{dx^\nu(\tau)}{d\tau}}\theta(t-t_p) \label{partac} \\
& & -\sum_s\int d^4 x \int d^2\sigma \delta^{(4)}(x^\mu-x_s^\mu(\sigma)) T(\phi(t,\bx))\sqrt{-Det g_{\mu\nu}(x_s(\sigma))\partial_\alpha x^\mu(\sigma) \partial_\beta x^\nu(\tau)}\theta(t-t_s) \nonumber
\eea
where $p$ indexes particle sources and $s$ indexes string sources with a mass $m$ or tension $T$ which in general depend on the inflaton field $\phi$.  The step functions in (\ref{actionfull}) reflect the fact that the particle or string sources are produced during inflation, on a timescale short compared to the scale of gravitational waves we wish to consider.

Here ${\cal S}_{XY}$ describes couplings of the $X$ sector (particles or strings) to other fields $Y$, included because this is a generic possibility which affects their decays.
Of course the $X$ sector necessarily couples to gravity, and in addition to producing GWs classically, $X$ strings can decay into gravitons, and two or more $X$ particles can annihilate into gravitons.

Clearly, solving for the detailed dynamics and GW spectrum from (\ref{actionfull}) is prohibitively difficult in general since the system is nonlinear; but this is also not necessary for our goal of estimating the leading contributions in some cases.  For sufficiently large density, for example, one may instead treat the collection of sources $X$ as a fluid, along the lines recently reviewed and applied in \cite{Caprini:2009fx}.  For sufficiently small density, it is tractable to sum the effects of individual sources.

\subsection{Examples of competitive effects}

Let us now consider some illustrative cases where the GW signature is competitive with (or exceeds) the tensor modes arising from the standard mechanism (\ref{inflGW}).  We will consider the production event itself, the effect of decays of the produced particles, and the effect of ordinary Bremsstrahlung radiation from scattering events.  Formally, these can all be thought of as Bremsstrahlung, putting appropriate incoming and outgoing lines in (\ref{BremGW}) (or a generalization of that equation to account for time dependent masses).

An individual production, scattering, or decay event produces gravitational waves at all frequencies below the inverse timescale of the event.  Multiple events, involving the same or different sources, can enhance the effect on the one hand, but also can introduce interference which suppresses the effect.  For example, in a gas of particles with an approximately spherically symmetric distribution of particle positions and velocities, the quadrupole vanishes to first approximation and the net GW spectrum is a subleading effect.  Moreover, if one considers the contribution of a single particle subject to multiple scattering events without relevant momentum loss, its net GW emission at very long wavelengths is simply determined by the scattering angle between the first and last event, with no enhancement from the additional events.  This follows from the positive and negative contributions in (\ref{BremGW}) for incoming and outgoing lines.  We will take these suppression factors into account in our estimates below.

\subsubsection{GWs from Production}\label{subsubsec:GWprod}

In this section, we discuss conditions under which the production event itself contributes a competitive tensor signal. Let us first consider gravitational waves, and then we will also address scalar emission.  To analyze this, we need to specify the functional form of $m(\phi)$.  During the production itself, we assume a coupling of the form $\phi^2\chi^2$, so that $m(\phi)=\phi\approx \dot\phi t$.  We will consider two examples for the later evolution:  (i) $m(\phi)$ continues to depend linearly on $\phi\approx \dot\phi t$, and (ii) $m(\phi)$ transitions to a constant at some time $t_c < H^{-1}$ after the production event.  In the appendix \S\ref{subsec:massfunctions}\ we describe a mechanism by which such a transition may arise.
In a production event, the homogeneous rolling scalar field loses energy into $\chi$ particles, and also into scalar radiation (reducing $\dot\phi$ in the process).
\vspace{0.3cm}

\noindent {\it Case (i): $m(\phi) =\phi\approx \dot\phi t$}
\vspace{0.3cm}

Case (i) is interesting, as it is in some sense simplest to consider the $\phi^2\chi^2$ model without assuming a more complicated functional form for the $\chi$ mass.  We will consider the GW emission arising from the sudden appearance of the produced particles and the associated scalar radiation modes.

The stress-energy tensor for the particles, obtained by varying (\ref{partac}) with respect to the metric takes the form
\beq\label{stresspart}
T^{\mu\nu}_{part} =\sum_n \delta^{(3)}(\vec x-{\vec x}_n(t)) \frac{p_n^\mu p_n^\nu}{p_n^0} \theta(t)
\eeq
where
\beq\label{momenta}
p^0= \frac{g\phi(t,\vec x)}{\sqrt{1-\dot{\vec x}^2}} ~~~~ \vec p = \frac{g\phi(t,\vec x) \dot{\vec x}}{\sqrt{1-\dot{\vec x}^2}}
\eeq
The spatial components are of the form
\beq\label{Tspatial}
T^{ij}\sim \frac{p^ip^j}{p^0}\delta(\vec x -{\vec x}_n(t))\theta(t)\sim \frac{p^ip^j}{g\dot\phi t}\delta(\vec x -{\vec x}_n(t))\theta(t)
\eeq
The additional $t$-dependence in the denominator in (\ref{Tspatial}) translates into an additional factor of $\omega$ in its Fourier transform relative to the case studied in \cite{Weinberg}\ which had a constant mass for the particles.  The resulting gravitational wave emission is of the order
\beq\label{dEdwtdep}
\frac{dE_{GW}}{d\omega}\sim \left(\frac{E}{M_P}\right)^2\left(\frac{\omega}{E}\right)^2
\eeq
(This scales like the result \cite{Weinberg}\ (\ref{BremGW}), times the extra factor of $(\omega/E)^2$ just noted.)

From this we obtain at frequencies of order $H$ that $\rho_{GW}\sim H n_\chi H^2/M_P^2$ where $n_\chi$ is the number density of produced particles and hence at frequencies $\omega\sim H$ we have
\beq\label{hprodtmass}
h^2\sim \frac{\rho_{GW}}{\rho_{total}}\sim f \frac{H^3}{E M_P^2}
\eeq
where we defined $f\equiv En_\chi/\rho_{total}$ (with $\rho_{total}\sim H^2M_P^2$) to be the fraction of the total energy density that is contained in the produced particles~\footnote{In deriving this formula, we have neglected the backreaction of the particle creation events on the inflaton. This is justified as long as the energy density in these particles is at most of the order of the kinetic energy of the inflaton. When backreaction is important, one needs to follow the same treatment as in~\cite{trapped}.  Notice that even though the mass of these particles increases linearly with time after their creation, the energy density stored in them saturates at the energy they have after one Hubble time. Indeed, if $a_i$ is the scale factor at the creation time $t_i$, when $n_0$ particles are produced, at generic time $t$ we have:
\be
\rho_\chi(t)\sim \sum_{i } n_0 \left(\frac{a_i}{a}\right)^3 \frac{\dot\phi}{H}\left(1+ H(t-t_i)\right)\sim \int^t_{-\infty} \frac{dt'}{\Delta t}\;  n_0\, e^{-3 H(t-t')} \frac{\dot\phi}{H}\left(1+ H(t-t')\right)\sim n_0\frac{\dot \phi}{H} \frac{1}{ H\Delta t}\ ,
\ee
where $\Delta t$ is the spacing in time between the production events. For us $n_\chi\sim n_0/(H\Delta t)$.}. As we explain more in detail in the next subsection, there is only a factor of $n_\chi$ here because all of the production events are independent.
Turning this around, we see that
\beq
\left(\frac{H}{M_P}\right)^2\sim h^2\left(\frac{E}{fH}\right)\ge 10^{-12}\times \left(\frac{E}{fH}\right)
\eeq
where in the last step we put in the condition that $h$ be detectible.  The second factor here is $> 1$, so the inflationary scale $H$ would have to be larger than $10^{-6}M_P$ in order for these particle sources to produce a detectible tensor signal.  But this would require a higher scale of inflation than gives a standard contribution (\ref{inflGW}) to tensor modes of order $h\sim 10^{-6}$.

However, there is also stress-energy in the inflaton field $\phi$, which can source GWs.  We can estimate the scalar radiation as in the appendix, focusing on the upper limit of the integral in (\ref{scalarpert}).  This yields an energy density of order
\beq\label{phien}
\rho_{\delta\phi}\sim \sqrt{\dot\phi} N_\chi H^3
\eeq
(at coupling $g\sim 1$),
with the $\delta\phi$ particles carrying typical energies of order $E_{\delta\phi}\sim\sqrt{\dot\phi}$.
(The latter follows because we choose the upper range of $k$ in (\ref{eq:deltaphi2}), since we get the largest GW signal and the largest energy denstiy from the largest energy $\delta\phi$ particles.)

From these $\delta\phi$ particles, we obtain GW Bremsstrahlung which is a factor of $E_{\delta\phi}^2/H^2$ times (\ref{hprodtmass}):
\beq\label{scalarGWprod}
h^2\sim f_{\delta\phi}\frac{H E_{\delta\phi}}{M_P^2}
\eeq
Using that, with now $E_{\delta\phi}\sim\sqrt{\dot\phi}\sim \epsilon^{1/4} (H M_P)^{1/2}$ we get:
\beq\label{otherway}
\frac{H}{M_P}\sim h\left(\frac{h^{1/3}}{ f_{\delta\phi}^{2/3}\epsilon^{1/6}}\right)\sim h \left(\frac{h^{1/3}}{ f^{2/3}\epsilon^{1/6}}\right)\times \epsilon^{1/6}\left(\frac{M_P}{H}\right)^{1/3}
\eeq
Here $f_{\delta\phi}$ is the fraction of the total energy density which is carried by the $\delta\phi$ particles.  This fraction is less than $f$, the fraction carried by the $\chi$ particles, since those are ramping up in mass throughout a Hubble time, ultimately reaching energy $E_\chi\sim \dot\phi/H$.  At $g\sim 1$ we have from (\ref{phien}) that $f_{\delta\phi}\sim H f/\sqrt{\dot\phi}$.  This was used in the last step of (\ref{otherway}).
Since $H/M_P\le h$, and since we require $f<\epsilon\ll 1$, we see from this that the signal is still too weak to be competitive
in this example.

Although this example with a simple $\phi^2\chi^2$ coupling does not work at the level of production by itself, we will find below in \S\ref{subsec:stringring}\ that subsequent decays can produce a very competitive signal.  Before considering decays, however, let us consider a second model for $m(\phi)$.
\vspace{0.3cm}

\noindent {\it Case (ii):  $m\to const$}
\vspace{0.3cm}

As our next example, let us consider case (ii) in which the mass becomes constant after a time $t_c\ll H^{-1}$, and does not interact further in a Hubble time.
In this case, for gravitational waves of frequency $\omega\sim H$, we can work directly with the results (\ref{BremGW}) from the time-independent analysis of \cite{Weinberg}.  This gives
\begin{equation}
h^2 \sim f {E H \over M_P^2} \frac{E}{M}\ .
\end{equation}
Here $E$ is the energy of the particle that is relevant for gravitational emission, while $M$ is the final mass of the particle.
As above, we are interested in comparing the amplitude of gravitational waves produced by Bremsstrahlung to those produced by ordinary inflation (\ref{inflGW}), and so let us rewrite this as
\begin{equation}\label{HMpii}
{H \over M_P}\sim h\  \left({h M_P \over f E} \frac{M}{E}\right).
\end{equation}
This means that for a given $h$ the value of $H/ M_P$ needed to produce it is a factor of
\begin{equation}
 {h M_P \over f E}  \frac{M}{E}\ .
\end{equation}
smaller than the standard value. Since $E < M_P$, $E< M$, and $f < 1$ we conclude it is only possible to reduce the Hubble parameter by an amount less than $h$ which at the limit of detectability is roughly $10^{-6}$.
Thus we have six orders of magnitude of potential gain.  To obtain the full six orders requires $E\sim M$, having particles of energy close to $M_P$  having energy density comparable to that in the inflaton, but simply to obtain a competitive signal requires
\beq\label{basicreq}
\frac{E^2}{M_P M} \ge \frac{h}{f}
\eeq
Particles are produced moderately relativistic at an adiabatic time $t_a\sim g^{-1/2} \dot\phi^{-1/2}$, and then they grow in mass up to time $t_c$ becoming in the meantime non-relativistic. In order to avoid a quadrupole suppression in the gravitational production, we will take $E\sim g \dot \phi t_a$, while the asymptotic value of their mass depends on $t_c$:
\beq\label{tcE}
M\sim g \dot\phi t_c,
\eeq
or equivalently
\beq\label{eq:ener}
E^2\sim M^2 \left( \frac{t_a}{t_c}\right)^2\sim \epsilon g^2 M_P^2 (H t_c)^2 \left( \frac{t_a}{t_c}\right)^2\ .
\eeq

Next we will study the consistency of the requirement (\ref{basicreq}) with the constraint imposed by the need to limit the scalar power emitted from the production event.  The extent of the coupling to the scalar is determined by $\partial_\phi m(\phi)$.  Once the mass becomes constant, there is no longer any coupling to the scalar, but during the period between $t=0$ and $t=t_c$ in which the mass grows linearly with $\phi$ (and with $t$), there is a constant coupling to the scalar and we wish to estimate the scalar power from this whole process.

For frequencies $\omega \ll 1/t_c$, there is destructive interference between the production event at $t=0$ and the event at $t=t_c$ when the coupling jumps to zero.  This cancelation is exact at $\omega=0$, but for nonzero $\omega$ there is a residual contribution that arises from expanding a factor of $\e^{i\omega t}$ that arises in the Fourier expansion of the radiation.  In the radiated power, this introduces a suppression factor of $(\omega t_c)^2$ relative to the case analyzed in (\ref{scalarpmat}) where there was no time $t_c$ at which the scalars decouple.  Because of the redshifting of modes within the Hubble patch before they freeze out, we get the largest contribution by taking modes of $\omega\sim H$ and paying this $(H t_c)^2$ suppression price.
The scalar fluctuations are of order:
\beq\label{Pzetaprodii}
\zeta^2\sim \frac{g^2 f}{\epsilon} \frac{H}{E} (H t_c)^2.
\eeq
To assess the viability of the scenario we compute the scalar to tensor ratio,
\beq
\frac{\zeta^2}{h^2}\sim \frac{g^2 f}{\epsilon} \frac{H}{E} (H t_c)^2\times  \frac{M_P^2}{f E H}\frac{M}{E}\sim \frac{1}{\epsilon^2} \left( \frac{t_a}{t_c}\right)^3\ .
\eeq
Thus this scenario is viable as long as $\epsilon\sim 10^{-1}$.

Since the new source for $\zeta$ fluctuations dominates, we need to ensure that the resulting scalar fluctuation satisfies the current bound on non-Gaussianities, at the level of $10^{-3}$. The amount of non-Gaussianity scales as $1/\sqrt{N_{\phi}}$, with $N_{\phi}$ representing the number of $\delta\phi$ fluctuations contained in an Hubble patch, and therefore we have the constraint $N_\phi\gtrsim 10^{6}$. The constraint on the power spectrum gives an upper bound to the value of $N_\phi$, and we need to check that there is an open window. We have
\be\label{eq:scalar_ng}
\frac{h^2}{\epsilon^2}\sim \zeta^2\sim N_\phi \frac{H^4}{\dot\phi^2}\quad\Rightarrow\quad N_\phi\sim \frac{g^2 f^2(H t_c)^2}{h^2}\left( \frac{t_a}{t_c}\right)\ ,
\ee
after using $\dot\phi^2\sim \epsilon H^2\mpl^2$ and (\ref{HMpii}) and (\ref{eq:ener}) to substitute for $(H/\mpl)$. We see that for extreme values $Ht_c\sim 1$, $g\sim 1$, $f\sim 10^{-1}$, $ t_a\sim t_c$ and $h\sim 10^{-6}$, $N_\phi$ can be as large as $10^{10}$, which gives quite a large window where the non-Gaussianity of the scalar fluctuations is compatible with observations, though this allowed window shrinks if we move ourselves away from the most extreme region of parameter space. Interesting there is also a non negligible region of parameter space where the resulting non-Gaussian signal is detectable, but not yet ruled out. If all these constraints are satisfied, we can obtain a dominant tensor contribution from our sources, with $H/M_P$ low enough to suppress the usual contribution (\ref{inflGW}).

Finally, let us check that the scattering rate $\Gamma$ is indeed $\le H$, since otherwise we would need to take into account additional interactions, and that $t_c\lesssim H^{-1}$.  The latter condition is (setting $g\sim 1$ for simplicity)
\beq\label{tccond}
{ M} \ll \frac{\dot\phi}{H}
\eeq
Regarding the scattering rate $\Gamma$, we have
\beq\label{Gammaii}
\Gamma\sim \sigma n_p v\sim \frac{1}{M^2}\times \dot\phi^{3/2}\times \frac{\sqrt{\dot\phi}}{M}\sim \frac{\dot\phi^2}{M^3}
\eeq
The last factor $v$, the particle velocity, is $p/\sqrt{p^2+m(t)^2}\sim p/m(t)$, where $p\sim\sqrt{\dot\phi}$ is the momentum of the created particles.  Now putting in (\ref{tccond}) we see that
\beq\label{Gammaok}
\Gamma \gg H \times \frac{H^2}{\dot\phi}.
\eeq
Since the latter factor is $\ll 1$ in general in our model, this is consistent with a sufficiently slow decay rate, $\Gamma \le H$.

\subsubsection{Decay of massive particles into massless ones}
\label{subsubsec:decaypart}

In this section, we consider gravitational waves produced during decays of massive particles $\chi$ present during inflation.  As we have discussed, such particles may be produced via a coupling to the rolling inflaton such as $g^2\phi^2\chi^2$; in some circumstances \cite{trapped}, this process repeats periodically during inflation.
Here we assume that $\chi$ couples to other light degrees of freedom $Y$, such that they can decay within a Hubble timescale.  Similarly to what happens  for the case of electromagnetic radiation in $\beta$-decay, gravitational Bremsstrahlung radiation is produced not only in the case of scattering of particles, but also in a decay process.

As we described above in \S\ref{subsec:generalsetup}, the amount of gravitational radiation per unit solid angle per unit frequency is given by \cite{Weinberg}
\begin{equation}\label{gwenergy}
\left( {dE \over d\omega d\Omega}\right) = {\omega^2 \over 2 \pi^2 M_P^2}\sum_{D_i,D_j} e^{i k_\mu (x_{D_i}^\mu - x_{D_j}^\mu)}  F[\{P_{D_i}\},\{P_{D_j}\},k]\ ,
\end{equation}
where
\begin{equation} \label{eq:factorF}
F[\{P_N\},\{P_M\},k]=\sum_{N,M} {\eta_N \eta_M \over P_N\cdot k \ P_M\cdot k} [(P_N\cdot P_M)^2 -{1\over 2} m_N^2 m_M^2]\ .
\end{equation}
Here $D_i$ labels the collisions, $x_{D_i}^\mu$ the location in space time where each collision/decay occurs, and $\{P_{D_i}\}$ is the set of momenta involved is each collision. $\eta$ is $-1$ for incoming particles and 1 for outgoing ones.

For a single decay of one massive particle with mass $M$ into two massless ones the above formula gives:
\begin{equation}\label{decayBrem}
\left( {dE \over d\omega d\Omega}\right) = {1\over 4 \pi^2} {M^2 \over M_P^2}\ .
\end{equation}
Here $(E/M_{pl})^2$ is the effective dimensionless gravitational coupling squared, analogous to $\alpha$ in electromagnetic Bremsstrahlung.

For an arbitrary  number of decays, the above formula becomes very complicated: events at different locations interfere if their relative distance is smaller than the frequency of the emitted wave, and the factor~(\ref{eq:factorF}) involves momenta that belong to different events.  In the present work, our main purpose is not to perform a precise calculation of the emitted gravitational radiation in one particular scenario. Most humbly we simply wish to estimate the order of magnitude of the effect in reasonable examples, in order to determine whether or not there is a direct relation between inflationary tensor modes and the height of the inflaton potential.

Tensor modes at wavelengths much shorter than $H^{-1}$ are highly suppressed by redshift (\ref{energetics}).  In each scattering, decay, or production event the energy produced in gravitational waves is frequency-independent for wavelengths longer than the timescale of the process.  For these reasons, the leading contribution comes from those gravitational waves that are produced directly with Hubble frequency.  Focusing on these wavelengths allows us to neglect the modulation due to the different location of the events, and consider them all at the same point in space.

We must take into account possible destructive interference among the various decays, which result from the sum over momenta at different events in~(\ref{eq:factorF}). The energy emitted in gravitational radiation goes as the square of the stress-energy tensor. So if there are $N_{\rm part}$ particles decaying in an Hubble patch, then naively the amount of gravitational radiation should go as $N_{\rm part}^2$. Clearly this is an overestimate. If these particles are randomly distributed in an Hubble patch, their decay products will tend to be spherically distributed in the limit in which $N_{\rm part}$ is very large, and this will lead to a suppression of the emission of gravitational radiation.
We can determine the net effect of the increased number of particles in the following way.
The quadrupole-squared of the random distribution of particles will have a typical size proportional to $N_{\rm part}$ instead of to $N_{\rm part}^2$. This is very much like the variance of $N$ independent random variables, which goes as $N$ and not as $N^2$, and also much like the typical realization of a random walk in quadrupole space, where we sum randomly all the quadrupoles associated to each event. Another way to think about this is to notice that the momentum sum in (\ref{eq:factorF}) assuming the two set of momenta belong to two different decays is not zero. However, it becomes zero if one averages over the outgoing direction of the momenta of the second decay. Now, in the case in which there are many decays happening at the same point, then the sum in (\ref{eq:factorF}) where the second momenta are taken from the various collisions effectively corresponds to averaging over the outgoing direction of the momenta of the second decay. In this case therefore the two sums over decays collapse to one, leaving us with a single factor of $N_{\rm part}$.

Altogether, if there are $N_{\rm part}$ particles decaying within an Hubble patch, with $N_{\rm part}$ large, the amount of produced gravitational radiation goes as
\begin{equation}
\left( {dE \over d\omega d\Omega}\right) \sim {1\over 4 \pi^2} {M^2 \over  M_P^2}\;\times\; N_{\rm part} \ .
\end{equation}
This leads to the following amplitude in gravitational waves
\begin{equation}
h^2\sim {1\over \rho_{total}} \left.{d\rho_{gw} \over d\ln \omega}\right|_{\omega\sim H}\sim {H \over \rho_{total}} \frac{ M^2 n_{\rm part}}{M_P^2}\ ,
\end{equation}
where $n_{\rm part}$ is the number density of the decaying particles before they decay and where we have neglected numerical factors. If we call $ f_{\rm }$ the fraction of the energy density carried by the particles prior to their decay, we can re-write the above expression in a more useful way
\be
h^2\sim  \frac{f H M }{\mpl^2} \ .
\ee
Let us see how big this number can be.

Since the decaying particles do not redshift as approximately a cosmological constant, we need to have $f_{\rm}\lesssim \epsilon$, with $\epsilon$ being the slow roll parameter, a number much smaller than one (but not necessarily tiny, let us say not larger than $10^{-1}$). Notice that for standard slow-roll inflation, saturating this bound means that a fraction of order one of all the kinetic energy of the inflaton is dissipated in the creation of particles. In more general models this does not need to be the case, but this is an interesting regime to consider since it provides a mechanism to dissipate excessive kinetic energy on a steep potential, as in trapped inflation~\cite{trapped}.

The mass $M$ of the decaying particle can be bound to be at most of order $M_{\rm pl}$. We therefore can write the above formula as
\be
h^2\sim \epsilon \frac{H}{\mpl}\cdot \frac{M}{\mpl}\cdot \frac{f_{\rm}}{\epsilon}\lesssim\epsilon\frac{H}{\mpl}\ .
\ee
Alternatively, we can express the necessary value of $H/\mpl$ in order to have detectable signal:
\be\label{HMP}
\frac{H}{\mpl}\sim h \left(\frac{ h \mpl}{f_{\rm} M}\right)\ ,
\ee
which implies that the value of $H/\mpl$ can be reduced from the standard case by a factor of order $h \mpl/(f_{\rm} M)$.
This is quite a big improvement with respect to the ordinary case $H/\mpl\sim h$. For example if we take $f$ of order $10^{-1}$ this means that we could detect gravitational waves for values of $H/\mpl$ as low as $10^{-11}$. This represents five orders of magnitude improvement with respect to the ordinary case (and ten orders of magnitude if you count in terms of the more physical parameter $H^2$).

With a large window of opportunity for competitive GWs, let us consider more specific examples, relaxing some of the assumptions just made.  For example,
rather than taking $M\sim M_P$, we can consider the mass which is built up after the specific production mechanism in the appendix, arising from a coupling of the form $g^2\phi^2\chi^2$, and check how massive the $\chi$ particles become before decaying.  If we call $\Delta t$ the time since the particle was created we can write:
\beq\label{chiphimodel}
M \sim g \dot\phi \Delta t \sim g\sqrt{\epsilon} M_P (H\Delta t).
\eeq
We can take $(H\Delta t)\sim 1$ and get:
\beq
\frac{H}{\mpl}\sim h \left(\frac{ h}{g f \epsilon^{1/2}}\right)\ .
\eeq
If we consider $f\sim\epsilon$ we  obtain that $H/M_P\sim 10^{-12}/g\epsilon^{3/2}$.  If $\epsilon\sim 10^{-1}$ and $g\sim 1$, we obtain $H/M_P\sim 3 \times 10^{-10}$, so the usual mechanism for tensor modes is suppressed by a large factor relative to the new sources in this example.

We can be a bit more general and relax the $(H\Delta t)\sim 1$ assumption. We can assume that there are many production events in a Hubble time, at a rate $dN_{hits}/dt$ and that in each of these production events $n_c$ is the number density of created $\chi$ particles and that the decay rate of the  particles is $\Gamma_d$. The energy density in the $\chi$ particles is given by:
\beq\label{rhochione}
\rho_\chi = \int d\Delta t \ n_c \frac{dN_{hits}}{dt} e^{-(3 H + \Gamma_d)\Delta t} M\sim n_c \frac{dN_{hits}}{dt} \tilde M \tilde t.
\eeq
with $ \tilde M \sim g\sqrt{\epsilon} M_P (H \tilde t)$ and  $\tilde t \sim{\rm min}[1/H,1/\Gamma_d]$. We have also replaced the discrete sum over production events by an integral. We conclude that the typical mass of the $\chi$ particles is determined by the shortest between a Hubble time and the lifetime because even if the lifetime is much longer than Hubble the abundance of very old particles dilutes exponentially. Only production events in the last $\tilde t$ contribute particles at any given moment because particles from previous events have either decayed or diluted.

We need to demand that the energy in the $\chi$ particles and its decay products ($Y$) be a small fraction of the vacuum energy driving the expansion. We must take into consideration that the energy in the decay products of $\chi$ does not have time to redshift during a Hubble time. This is especially relevant when $\Gamma_d >> H$. We have
\beq
\rho_\chi + \rho_Y \sim  f H^2 M_P^2 \sim  n_c \frac{dN_{hits}}{dt} \tilde M \frac{1}{H},
\eeq
which is just an energy $\tilde M$ for each of the particles created in a Hubble time wether they have decayed or not.

The amplitude of the tensor modes can be estimated by adding the contributions from all the $\chi$ decays in a Hubble time. For simplicity we will approximate the density of $\chi$ particles as constant during a Hubble time, given by $\rho_\chi$ in equation  (\ref{rhochione}) divided by $\tilde M$. We then have:
\beq
h^2 \sim  \frac{f H \tilde M}{M_P^2}  \times
\left\{
\begin{array}{ll}
\frac{\Gamma_d}{H}\  , & {\rm if} \ \  \Gamma_d < H  \\
1\  , & {\rm if} \ \  \Gamma_d > H \ ,
\end{array}
\right.
\eeq
where the factor $\Gamma_d/H$ accounts for all the decays in a Hubble time. Thus for a fixed $f$ the tensor amplitude is maximized when $\Gamma_d \sim H$. For $\Gamma_d << H$ there are too few decays in a Hubble time to produce a lot of gravity waves and if $\Gamma_d >> H$ the mass of the $\chi$ particles at the time of their decay is not large. We can turn this into an estimate for the Hubble scale for a given $h$:
\beq
\frac{H}{M_P}\sim  h \left(\frac{ h}{g f \epsilon^{1/2}}\right)
\times
 \left\{\begin{array}{ll}\left(\frac{H}{\Gamma_d}\right)\  , & {\rm if} \ \  \Gamma_d < H \;; \\ \left(\frac{\Gamma_d}{H}\right), & {\rm if} \ \  \Gamma_d > H \,. \end{array}\right.
\eeq
This is the same result we got when assuming $H\Delta t\sim 1$ but enhanced by a factor $H/\Gamma_d$ or $\Gamma_d/H$ depending respectively on wether $H$ is larger or smaller than $\Gamma_d$.

As before we need to make sure that the scalar power is below or at the same level as observed. The estimate is very similar to the example of gravity waves created at production which we analyzed before, in particular the case when the scalar coupling turned off. Again we could imagine than now the scalar coupling turns off prior to the decay, at a time $t_c$. If this is so the mass of a $\chi$ particle  does not continue to grow after $t_c$ and thus $t_c$ provides an upper limit for $\tilde t$, $\tilde t \rightarrow {\rm min}[\tilde t, t_c]$. As before the scalar power is suppressed by a factor $(H t_c)^2$. For the purpose of this suppression, the case when the scalar coupling never turns off corresponds to $t_c\sim \tilde t$ as the decay of the $\chi$ into particles with no scalar charge effectively acts in the same way as the shutting off of the scalar coupling and the Hubble time  also provides a bound to the possible level of cancelations. Thus we always have $t_c\sim \tilde t$.

We can now compute the ratio between scalar and tensor power,
\beq
\frac{\zeta^2}{h^2}\sim \frac{g^2}{(\tilde M /M_P)^2} \times (H t_c)^2\times  \frac{H^2 M_P^2}{\dot \phi^2}\sim \frac{1}{\epsilon^2},
\eeq
where the first term accounts for the ratio of couplings, the second for the suppression of Bremsstrahlung in the scalar case and the third comes from the conversion between $\phi$ and $\zeta$ fluctuations and the normalization of the gravity wave energy density. Thus the scenario is viable as long as $\epsilon\sim 10^{-1}$. Following the same steps that led to eq.~(\ref{eq:scalar_ng}), it is straightforward to check that the scalar power spectrum can easily satisfy the constraint on non-Gaussianity.

Finally we need to check wether it is possible to neglect the annihilation of $\chi$ particles into $\delta\phi$ particles. The rate for this reaction goes like $\Gamma\sim \sigma n v\sim \sigma N_{\rm part}H^3 v$ (where $v$ is their velocity and $\sigma$ their annihilation cross section).
The cross section for $2\to 2$ scattering from the interaction term $g^2\phi^2\chi^2$ goes like
\beq\label{cross}
\sigma\simeq \frac{g^4}{(8\pi)^2}\frac{1}{E_\chi^2} \lesssim \frac{g^4}{(8\pi)^2}\frac{1}{\dot\phi}
\eeq
where in the last step we used that the energy $E_\chi$ carried by the massive $\chi$ particles is $\ge \sqrt{\dot\phi}$  (it goes like $M\gg\sqrt{\dot\phi}$ at late times).     Multiplying this through by $N_{\rm part}H^3 v$, and including a factor $N_{hits}$ for the number of production events during a Hubble time, we get
\beq\label{decayrate}
\Gamma\lesssim H\left(\frac{g^3}{(8\pi)^2} N_{hits}\right)
\eeq
This is less than $H$, and hence completely negligible, as long as $N_{hits}<(8\pi)^2/g^3$, a condition which is easy to satisfy.

\subsubsection{Creation of string pairs in the case that they decay into rings of particles}\label{subsec:stringring}

Next, let us consider a microscopic example leading to decay-induced Bremsstrahlung gravitational radiation.  In this example, we consider a situation where the rolling inflaton first produces pairs of strings, which decay into smaller string loops and then into particles.
Consider a pair of long strings of length $L$ produced via a time-dependent string tension, $T\approx \dot T t$.  In the appendix below, we discuss the production of pairs of these strings, finding that the two members of the pair are created close to each other and with opposite orientation.   The system may produce loops which oscillate relative to each other on a timescale of order $\dot T^{-1/3}$, although it may also produce straighter stretched strings on a shorter timescale.  In this work we will focus on the former case.
These  relatively oscillating strings can quickly (on a timescale of $\dot T^{-1/3}\ll H^{-1}$) fall apart into $L {\dot T}^{1/3}$ smaller string loops.\footnote{We thank J. Polchinski for this point.}  (Here we conservatively assume there are no small couplings in the system that suppress the interconnection of the strings; this is a feature of the ``tensionless string theories" which arise on branes in the relevant string constructions.)
The total angular momentum of the original pair of long strings is zero, and the small loops into which it breaks do not have any preferred orientation or direction of spin; we will therefore treat these as random in our estimates.
These smaller loops can then decay, in particular into scalar modes (and other light particles such as gravitons), with a random distribution of momentum directions.  There is no quantum number protecting them from decay, and no small coupling which suppresses their decay rate.

The original pair of long strings is not spherically symmetric, and supports a quadrupole. However, given the breakup into smaller loops and decay into light particles, for the purpose of long-wavelength emission we may describe the system as an instantaneous production of the final decay products.     The final decay products, for example $\delta\phi$ perturbations, do not have time-dependent masses, and so the analysis of \cite{Weinberg}\ goes through unmodified.

Including this, in the same way as in \S\ref{subsubsec:decaypart}\ we obtain
\beq\label{scalingStr}
\frac{dE}{d\omega d\Omega}\sim \frac{\dot T^{2/3}}{M_P^2}N_{loops}N_{rings}
\eeq
where $N_{rings}$ is the number of produced pairs of long strings (which then decay into rings of particles)
Hence, using $N_{loops}\sim {\dot T}^{1/3}/H$,
\beq\label{GWestStr}
\rho_{GW}|_{\omega\sim H}\sim H \times H^3\times \frac{\dot T^{2/3}}{M_P^2}\frac{{\dot T}^{1/3}}{H} \times N_{rings}\sim H^4 N_{rings}\frac{\dot T}{HM_P^2}
\eeq
leading to
\beq\label{GWhStr}
h^2\sim \frac{\rho_{GW}}{\rho_{Tot}}|_{\omega\sim H} \sim \left(\frac{H}{M_P}\right)^{2}N_{rings}\left(\frac{\dot T}{HM_P^2}\right)
\eeq

In order to assess the strength of this effect, we need a model which determines $\dot T$.  In the scenario \cite{monodromy}, $\dot T\sim \eta M_P \dot\phi$ with $\eta<1$ a coupling.  Since $\dot\phi/HM_P\sim\sqrt{\epsilon}\ll 1$, the GW emission from a single ring of loops is subdominant to the contribution from tensor quantum fluctuations, by a factor of $\eta\sqrt{\epsilon}\ll 1$.  To get a competitive or better signal, we require
\beq\label{Nrings}
N_{rings}>\frac{1}{\eta\sqrt{\epsilon}}
\eeq
We can bound $N_{rings}$ above by imposing that the total energy $N_{loops}N_{rings}{\dot T}^{1/3}H^3$ be less than $\epsilon H^2 M_P^2$ with $N_{rings}>HM_P^2/\dot T$ so as to produce a viable signal.  This leads altogether to the condition $\dot T^{1/3}>H/\epsilon$, which is easy to satisfy.

Finally, given the order one coupling of our strings to scalar perturbations $\delta\phi$, we must check if that would produce a contribution to the scalar power spectrum which is too large.\footnote{ We thank M. Mirbabayi for extensive discussions.} 
By a calculation similar to that in our appendix A.2.1, we find scalar emission during the production period before the long strings decay, of order
\be\label{zetastrings}
\zeta_{long}^2\sim \frac{1}{\epsilon}\eta^2 N_{rings}(Ht_c)^2
\ee
where $t_c\sim \dot T^{-1/3}$.
This leads to strong constraints on this case from bounds on scalar modes, with $h^2/\zeta^2\sim \epsilon^2/N_{loop}$; it is weaker than the particle decay case by the factor of $N_{loop}$ but could give marginally competitive results.  It remains to be seen what the contribution is from strings which are produced more quickly, without oscillations, and subsequently decay.  We leave this question for future work.

There are also $\delta\phi$ modes as decay products of the small loops.  These give a subdominant contribution to the scalar emission, as follows.
We expect that the energy goes into scalars with wavelength of order ${\dot T}^{-1/3}$ to first approximation.  In particular, decay into a large number of low-frequency modes is suppressed.  The system is weakly coupled for energies much less than the string tension, which at time $\dot T^{-1/3}$ is of order $\dot T^{2/3}$.  Emission into many $\delta\phi$ particles of low frequency $\omega\sim \dot T^{1/3}/n$ is suppressed by a factor $(\omega/\dot T^{1/3})^n\propto n^{-n}$.

Given that the strings decay preferentially into $\delta\phi$ particles of frequency $\sim \dot T^{1/3}$, the resulting contribution to their energy density at freezeout is
$\Delta\rho_{\delta\phi}\sim \epsilon H^2 M_P^2(H/{\dot T}^{1/3})^4$ (where the last factor accounts for the redshifting before freezeout).  The ratio of this to the usual source of scalar perturbations (for which $\rho_{\delta\phi}\sim H^4$) is $\epsilon H^2M_P^2/{\dot T}^{4/3}$.  Evaluating this for $\dot T^{1/3}\sim \xi H/\epsilon$ (with $\xi\ge 1$), we obtain  a ratio $\Delta\rho_{\delta\phi}/\rho_{\delta\phi}\sim \xi^{-4}\epsilon^5 M_P^2/H^2$.  For $\epsilon\sim 10^{-2}$, this is easily subdominant for modest values of $\xi$.

\subsubsection{Bremsstrahlung from Scattering}

The final case we will study is the Bremsstrahlung radiation produced is when particles accelerate due to collisions.
In a process where an energy $E$ is transferred, Bremsstrahlung at low frequencies $(\omega \ll E)$ emits an equal amount of energy per unit frequency
\begin{equation}\label{single}
{d E_g \over d \omega}\approx \left({E\over M_P}\right)^2.
\end{equation}
for a single scattering event.  However, in a gas of particles there will be multiple interactions.  At sufficiently low frequency, the emission might be suppressed if particles interact before they can emit a graviton. We will parameterize this as:
\begin{equation}\label{pcases}
{d E_g \over d \omega}\approx \left({E\over M_P}\right)^2 \left\{\begin{array}{ll}1\  , & {\rm if} \ \  \omega > \gamma^2\,\Gamma_{int}\;; \\ \left(\frac{\omega}{\gamma^2\,\Gamma_{int}}\right)^{p+1}, & {\rm otherwise\,.} \end{array}\right.
\end{equation}
Here $\Gamma_{int}$ is the rate of interaction in the frame in which our gas of particles overall has no net momentum and $\gamma$ is the boost factors of the scattering particles.

These suppression factors \cite{Koshelkin:2004as}\ can be intuitively understood as the following.\footnote{We thank M. Peskin for a useful discussion of this issue.}  Consider a process where a particle interacts twice, hitting two targets which are a distance $\ell$ apart (in our cosmological frame).  The rate of interaction is $\Gamma\sim v_\parallel/\ell$, where $v_\parallel=v\, \cos\theta$ is the component of the particle's velocity which is along the direction from the first to the second scattering event (in terms of an average scattering angle $\theta$).
The first interaction produces some GWs by Bremsstrahlung, and would give (\ref{single}) if that were all that happened. In the second interaction, the particle emits by Bremsstrahlung a second graviton.  This new wavefront interferes with the one emitted in the first interaction, effectively creating a higher frequency graviton.   The frequency of the higher frequency graviton is determined by how far the first emitted wavefront gets before the particle scatters again. From the wavefront geometry we have
\be\label{gammas}
\omega^{-1} = \ell/v_\parallel - \ell/c \simeq l /\gamma^2 v_\parallel\ ,
\ee
where in the second passage we have approximated $1-v_\parallel^2/c^2\sim 1-v^2/c^2=1/\gamma^2$ which is valid for not too large deviation angles and $v<c$.
We also have $\Gamma = v\, \cos\theta/\ell$, so we can rewrite this in terms
of $\Gamma$ instead of $\ell$.  This gives
\be
\omega \sim \Gamma \gamma^2\ .
\ee
For massless particles, $v=c$ and it is necessary to retain the $\theta$ dependence in (\ref{gammas}), giving
$\omega\sim \Gamma cos(\theta)/(1-cos(\theta))$.

For the moment we take the factor of $p$ as a free parameter, and will analyze the GW signal in various ranges of $p$.  We will comment later on its possible values in particular cases which have been analyzed in the literature. It should also be stressed, as we will highlight next, that the threshold frequency $\gamma^2\Gamma$ at which the suppression starts, is only the first of a series of thresholds in which different physical mechanisms suppressing Bremsstrahlung become important and the value of $p$ changes accordingly. We will see an example of this next. We take our initial threshold to be $\gamma^2\Gamma$, which is the one that occurs when suppression is due to multiple scattering, as a guidance which is particular relevant for our setup. Generalization to different setups should be straightforward.
\vspace{0.2cm}

Let us assume that there is a number density of particles $n_p$ each with typical energy $E$ interacting with a rate $\Gamma_{int}$. Since during inflation we expect the density of gravitational waves to be stationary, we then have:
\begin{equation}
{d\rho_{gw} \over d \omega} \sim n_p \frac{\Gamma_{int}}{H}  {d E_g \over d \omega}\ .
\end{equation}
Even without assumption of stationarity, because of redshift, we are interested only in gravitons produced in about an Hubble time. This leads to the same factor of $1/H$ above.
By using $n_p =\rho_p/ E$, and redshifting the gravitational waves down to the Hubble scale where they freeze out, we obtain:
\begin{equation}
h^2\sim {1\over \rho_{total}} \left({d\rho_{gw} \over d\ln \omega}\right)_{\omega\sim H} \sim f \times \frac{\Gamma_{int}}{H}  \times {\omega\over E} \times  {d E_g \over d \omega}\times \left({H \over \omega}\right)^4\ ,
\end{equation}
where we defined $f={\rho_p/ \rho_{total}}$.

We would like to understand if this is $\gtrsim 10^{-12}$ in a reasonable window of parameters, and whether this occurs in situations where the scale of inflation is too low to produce detectable tensor modes by the usual mechanism (\ref{inflGW}).

\paragraph{Thermal equilibrium} Since we are interested in gravitons produced in about an Hubble time, we need to have $\Gamma_{int}$ to be at least as large as $H$. Since $\Gamma_{int}$ is the numerator, let us start with a large $\Gamma_{int}\gg H$. This implies we are in thermal equilibrium.
We will include the possibility of a nontrivial species number $N_*$ and consider relativistic particles, giving
\begin{eqnarray}\label{eq:Gammaconstraint}
\rho_p  \sim  N_\ast T^4 \ , \qquad E  \sim &T \ ,\qquad \Gamma_{int} \sim   N_\ast \alpha^2 T\ ,
\end{eqnarray}
where $\alpha$ is the strength of the interaction.
We can use this to solve for $T$:
\begin{equation}\label{eq:Tconstraint}
 {T \over  M_P}\sim \left({f  \over  N_\ast}\right)^{1/4}\left({H  \over  M_P}\right)^{1/2} .
\end{equation}
The amount of gravity waves produced in this case becomes
\be
h^2\sim  f\frac{ H^3 T}{\mpl^2\gamma^6\Gamma_{int}^2}\left(\frac{\omega}{\gamma^2\,\Gamma_{int}}\right)^{p-2}\ .
\ee
We will have to take into consideration the suppression of the Bremsstrahlung emission. Depending on the strength of the suppression, the value of the index $p$, it will be advantageous to consider waves emitted at either $\omega\sim H$ or $\omega\sim\gamma^2\, \Gamma_{int}$. We consider each case in turn. 

\subparagraph{Index  $p\ge 2$:} In this case the suppression is strong enough that one is better off with GW with $\omega\sim \gamma^2\, \Gamma_{int}$. We then get:
\begin{equation}\label{eq:temp}
h^2 \sim  f  {H^3 T \over \Gamma_{int}^2  M_P^2 \gamma^6 }\ .
\end{equation}
Note that $\Gamma_{int}$ is in the denominator so that we are better off having the smallest possible rate compatible with thermal equilibrium $\Gamma_{int}\sim H$. Also, in order to minimize the redshift, it will be convenient to take $\gamma\sim 1$.
By using (\ref{eq:Tconstraint}), we can rewrite (\ref{eq:temp})  in terms of $H/\mpl$ to obtain:
\begin{equation}
\left({H \over  M_P}\right) \sim h \times  \left[\left({h^{1/3}\gamma^4 N_*^{1/6} \over  f ^{5/6} }\right) \left({\Gamma_{int}  \over  H}\right)^{4/3}\right].
\end{equation}
The factor in parenthesis on the right represent how much we can reduce the value of $H/\mpl$ with respect to the standard value $H/\mpl\sim h$.
Reducing $\Gamma_{int}/H$ to order one and by taking the extreme case $N_{*}\sim f\sim 1$ and $h\sim 10^{-6}$, we see that there are at most  two orders of magnitude to be gained and the window closes as the interaction rate increases above $H$. More realistically $f$ should be smaller than $10^{-1}$, decreasing the possible improvement to about one order of magnitude.

\subparagraph{Index  $0\le p\le 2$:} In this case the suppression is sufficiently weak that one is better off with GW with $\omega\sim H$. We have
 \begin{equation}
h^2 \sim  f \times \frac{ \Gamma_{int}^{-p} H^{p+1} T } {\gamma^{2p+2} M_P^2}.
\end{equation}
Note that here with  $0\le p\le 2$, $\gamma$ and $\Gamma_{int}$ are in the denominator, so one better off with the lowest possible $\Gamma_{int}\sim H$ consistent with thermal equilibrium and with $\gamma\sim 1$. We obtain:
 \begin{equation}\label{eq:generic_low_p2}
\left({H \over  M_P}\right) \sim h  \left({h^{1/3}N_*^{1/6}\over  f ^{5/6} }\right) \left({\Gamma_{int}  \over  H}\right)^{2 p/3} \gamma^{4(p+1)/3}\ ,
\end{equation}
which, after saturating the limit $\Gamma_{int}\sim H$ and minimizing by taking $\gamma\sim 1$, reduces to what we obtained the former case $p>2$, leading to at most two orders of magnitude in possible gain, even in the extreme case.

In the world of collider physics, detailed studies of the Bremsstrahlung suppression have been performed in the case of a relativistic particle impacting on a fixed target, as this is the case of relevance in particle detectors and in fixed target experiments. Here we follow the recent treatment of~\cite{Koshelkin:2004as}, where it is found that a value of $p=1/2$ can be obtained in the case of ultrarelativistic scattering for frequencies in the interval $\gamma^2\,\Gamma_{int}\gtrsim \omega\gtrsim \Gamma_{int}/\gamma^2$. For lower frequencies $\omega\lesssim \Gamma_{int}/\gamma^2$, the suppression factor $p$ becomes equal to $p=2$~\footnote{This is due to the fact that the small-angle approximation for the particle trajectory is no longer valid for such low frequency gravitons. We should point out that the minimum frequency at which we find Bremsstrahlung emission to be suppressed with a $p=1/2$, the so-called LPM regime, is different from the value quoted in~\cite{Koshelkin:2004as}. But our crossover scale, determined by the geometry of the scattering events described above, agrees with the one quoted in the rest of the  literature~\cite{Klein:1998du}.}.  Notice that we have a non-vanishing interval of frequencies where the suppression is controlled by $p=1/2$ only in the ultrarelativistic limit $\gamma\gg 1$, so we cannot take $\gamma\sim 1$ in~(\ref{eq:generic_low_p2}), which would increase the possible gain.  In this case (p=1/2), we have
 \begin{equation}\label{eq:gain_p_half}
\left({H \over  M_P}\right) \sim h  \left({h^{1/3}N_*^{1/6}\over  f ^{5/6} }\right)  \gamma^{2}\ ,
\end{equation}
If we impose the constraint  $\Gamma\gtrsim H\gtrsim \Gamma/\gamma^2$, then the gravitational modes $\omega\sim H$ which we are considering have frequency large enough to avoid the suppression $p=2$, which kicks in at $\omega\sim \Gamma/\gamma^2$, while still avoiding any suppression from redshifting. This constraint implies  that the gain is reduced to about one order of magnitude or a little more than that.

It is not hard to imagine setups in which during the inflationary epoch there might persist the same conditions as in colliders, at least in principle. In particular, we need particles to interact in a highly boosted regime so that the scattering angle is small, since in the opposite regime the work \cite{Koshelkin:2004as}\ argues that the suppression factor turns into $p=2$. As a proof of principle, let us take for example the case in which the inflaton produces two kind of species of particles due to its motion: one heavy non-relativistic scalar particle $H$ and one light relativistic scalar particle $L$. If the only interaction is a quartic vertex $HHLL$, then we see that these light particle will thermalize by scattering with the bath of $H$ particles, each interaction happening in the same kinematic regime as in cases studied for the particle colliders. $LL$ elastic scattering is suppressed as it happens only due to the mediation of a loop of $H$ particles, which can be thought to be very heavy with respect to the energy of the $L$ particles. The Bremsstrahlung radiation produced in the $LH$ interactions would then be expected to have a suppression factor $p=1/2$, based on the results of~\cite{Koshelkin:2004as}.

\paragraph{Particles with a few additional interactions:} So far in this section we have assumed that the particles were in thermal equilibrium, though sometimes we were led to take $\Gamma_{int}\sim H$. In this regime, it is not clear if the particles should obey a thermal distribution, and therefore, in order to explore the range of possibilities, it is interesting to consider the separate regime where the particle interact with a rate of order Hubble, but are not in thermal equilibrium.

Consider the case when after production the particles interact a few more times in a Hubble time. This will allow us to solve for the energy using that:
\begin{equation}
H\sim \Gamma_{int}  =  n_p \langle\sigma v\rangle  \sim n_p {\alpha^2 \over E^2},
\end{equation}
with $\alpha$ strength of the interaction. This together with $n_p E \sim f \rho_{total} \sim f H^2  M_P^2$ results in an expression for the typical energy:
\begin{equation}
 E \sim  M_P \left({f \alpha^2 H \over  M_P}\right)^{1/3},
\end{equation}
which leads to
\begin{equation}
 h^2  \sim \frac{f^{4/3}\alpha^{2/3}}{\gamma^6}\left(\frac{H}{\mpl}\right)^{4/3}\left(\frac{\omega}{H\gamma^2}\right)^{p-2}\ .
\end{equation}
Depending on the value of $p$, we are better off with emitting gravity waves at frequencies of order $\gamma^2\Gamma\sim\gamma^2 H$  ($p> 2$) or $H$ ($p\leq 2$).
In either case we obtain:
\begin{equation}
 \left({H \over  M_P}\right) \sim h  \left({h^{1/2} \gamma^q \over  f \alpha^{1/2} }\right) \ ,
\end{equation}
where $q=9/2$ for  $p>2$ or $q=3p/2$ for  $p\leq2$. We can reduce $H$ at most by three orders of magnitude.

\paragraph{Away from equilibrium:}
At then end of this section, we are naturally led to consider the case in which $\Gamma_{int} \lesssim H$ and thus we are not near thermal equilibrium.
The particles we consider are a result of the decay of the inflaton so they go through at least one interaction, when they are created. The Bremsstrahlung from their creation process was analyzed in detail above in \S\ref{subsubsec:GWprod}, and, as we saw, we obtained a larger window of GWs.
\vspace{0.2cm}

\paragraph{Scalar perturbations}

As in the previous examples, we must ensure that the scalar perturbations are consistent with current data and constraints.  In the present case of production of gravitational waves through scattering, the connection to the scalar fluctuations is model dependent. We have analyzed the Bremsstrahlung radiation in this section assuming that the scattering particles do not have any coupling to the inflation, as occurs for appropriate mass functions.
This reduces the question to whether the scalar modes from the production process are too great, in a case where the mass of the created particles becomes constant after production.  As discussed above (\ref{Pzetaprodii}), we have
\beq\label{Pzetaprodiiagain}
\zeta^2\sim \frac{g^2 f}{\epsilon} \frac{H}{E} (H t_c)^2
\eeq
Now $t_c$ is related to the mass $M$ ultimately attained by the created particles:
\beq\label{tcM}
H t_c \sim \frac{HM}{g\dot\phi}\sim \frac{1}{g\sqrt{\epsilon}}\frac{M}{M_p}\ .
\eeq
In the case in which the particles are created non-relativistically, scalar production  is minimized for $E\sim M$. This leads to
\beq\label{Pzetaprodmass}
\zeta^2\sim\frac{g f}{\epsilon^{3/2}}\left(\frac{H}{M_p}\right)(Ht_c)\sim \frac{ f}{\epsilon^2}\frac{H M}{M_p^2}
\eeq
which needs to be $\le 10^{-10}$.  In the examples in thermal equilibrium above, using (\ref{eq:Tconstraint}) we find
\beq\label{Tzeta}
\zeta^2\sim \frac{g f^{5/4}}{N_*^{1/4}\epsilon^2}\left(\frac{H}{M_p}\right)^{3/2}\lesssim 10^{-10}
\eeq
With an order of magnitude gain in our GW signal, $H/M_p\sim 10^{-7}$ and this marginally fits.
Similar comments apply to the case of particles with a few additional interactions.

\section{Discussion}

We have seen that a detection of tensor modes, even approximately scale-invariant tensor modes, from inflation does not automatically constitute a measurement of the inflationary potential energy.  We summarize our findings on the allowed values of $H$ for the various mechanisms is Table~\ref{Table:final}.
In this section, we discuss possibilities for distinguishing tensor modes from sources of the sort we have considered in this work from the standard source (\ref{inflGW}), in the event of a detection of primordial B modes.

\begin{table}

\begin{center}
    \begin{tabular}{ | l | l | l | p{5cm} |}
    \hline
    Kind of production & sub-Kind  & Approx. Max. Gain: $\log_{10}\left(\frac{H_{\rm min}}{H_{\rm min, \, vacuum}}\right)$ \\ \hline
     Particle Creation   & time-dependent mass & no gain \\  \hline
                    & time-constant mass & $\sim 4-5$,  or $\sim 2$ if very conservative\\ \hline
                            Decay  of particles   &  & $\sim 4-5$ \\  \hline
             Decay  of strings   & relatively oscillating & $\sim 1-2 $         \\ \hline
                       Scattering & Thermal Equilibrium &  $\sim 1-2$ \\ \hline
                       &  non-Thermal Equilibrium   & same as decay of particles \\ \hline
    \end{tabular}
\end{center}
\caption{\label{Table:final}\small Summary of the potential maximum gain in the value of $H$ with which we can obtain a detectable signal, according to the 3f production considered in the paper. }
\end{table}

\subsection{Power spectrum}

Our sources may be approximately scale invariant (e.g. in \cite{monodromy,trapped}) with the inflaton coupling to new sectors of light degrees of freedom which are closely spaced along its trajectory.  In that case, their power spectrum would be similar to that expected from (\ref{inflGW}).  However, if the inflaton couples to light fields more sporadically, one could obtain a non-scale-invariant signal, which would distinguish the sources from (\ref{inflGW}).  In the scale-invariant case, it would be interesting to consider additional methods for breaking the degeneracy.

One approach is to consider non-Gaussianity.  In the next subsection, we will discuss non-Gaussianity of the tensor spectrum. This was studied recently in \cite{Maldacena:2011nz}\ for vacuum fluctuations, where strong constraints on the shape were found.  In the presence of a source sector of the kind we have here we expect a wider range of possible shapes for the three point function, though we will only estimate their magnitude here.  It is worth noting that scalar non-Gaussianity arises in some of our examples:  if the energy in produced sources is large (of order $\epsilon H^2 M_P^2$), and the mechanism \cite{trapped}\ applies, then non-Gaussianity among the scalar modes is also predicted.

\subsection{A non-Gaussian window?}

In this section we have seen that there are several physical mechanism that could produce a signal in gravity waves that would be detectable and larger than the standard one due to the vacuum fluctuations. This means that we cannot derive the energy scale of inflation from observation of a scale invariant spectrum of gravity waves. It is worth pointing out that the scale invariance of gravity waves would still teach us about the time-translation invariance of the background when these modes were produced, a pristine signature of the quasi de-Sitter background which model-independently characterizes inflation ~\cite{EFT}. It is however interesting to ask if it will be possible to distinguish between gravity waves as due to our mechanisms (in the scale invariant case) and the standard ones due to vacuum fluctuations. A possible distinction might arise from observation of the statistical properties of the gravity waves. Though it would be interesting to make a comprehensive study of this phenomenon using the Effective Field Theory of Inflation~\cite{EFT}, it is hard to imagine that vacuum fluctuations of gravitons during inflation can be very non-Gaussian. This is because the energy scale controlling the free Lagrangian for the gravitons is $\mpl$, and this is a very high energy scale compared to the inflationary scale and to the usual canonical normalization of scalar fluctuations $\dot H^{1/2}\mpl/c_s$. Instead the new mechanisms of production that we have been describing in this section have nothing to do with vacuum fluctuations. Indeed, they are the result of interactions, and so are naively very non-Gaussian. However, there is another mechanism making to distribution Gaussian. This is the high number of gravitons being produced. Since each production event is independent, in the limit of high number of events the distribution becomes Gaussian, with a deviation of non-Gaussianity, parametrized by a dimensionless number that we call $NG$ and that can be thought of as the skewness of the distribution~\footnote{In terms of the usual parameter $f_{\rm NL}$, we have $NG\sim f_{\rm NL} h$}, that scales as the inverse square root of the number of gravitons at a given frequency at horizon crossing $N_{\rm gravitons}^{-1/2}$. It is easy to estimate that the number of graviton is given by
\be
N_{\rm gravitons}\sim H^{-4}\rho_{\rm gravitons}(\omega\sim H)\sim \frac{\mpl^2}{H^2} h^2 \ .
\ee
Notice indeed that this is order one for vacuum fluctuations. If we define the gain factor $g_{\rm gain}$ as a number in greater than one corresponding to the amount of decrease in $H/\mpl$ that we can have while still achieving a detectable signal with respect to the standard mechanism
\be
\frac{H}{\mpl}\sim \frac{h}{g_{\rm gain}}
\ee
we have
\be
g_{\rm gain}\sim \sqrt{N_{\rm gravitons}}\quad\Rightarrow \quad g_{\rm gain}\sim \frac{1}{NG}\ .
\ee
The amount of non-Gaussianity scales inversely to the gain. Observational constraints on $NG$ scale as the inverse square root of the number of modes that are signal dominated $N_{\rm modes}^{-1/2}$. Given the smallness of the signal in gravity waves, it is hard to imagine that  non-Gaussianities will be tested below the level $10^{-1}-10^{-2}$, implying that we would be able to distinguish these two scenarios only in the regime of modest gain, which is nevertheless a relevant fraction of the parameter space we have found with our examples. It is finally interesting to point out that there are also non-negligible regions of parameter space where the scalar fluctuations $\zeta$ present some detectable non-Gaussian features.  It would be interesting to study the details of the non-Gaussian distribution in the future.

\subsection*{Acknowledgements}
We would like to thank J. Polchinski for many useful discussions and collaboration on string production \cite{stringprod}, and D. Spergel for raising a question motivating part of this work. We thank Mehrdad Mirbabayi  for identifying three overestimates that appeared in the first version of this paper but that do not alter the conclusions:  one in the LPM formula in the scattering production mechanism, one related to subtlety in the gravity waves emitted from the particle creation, and a last one related to scalar production in the string case. We are grateful to L. McAllister, K-W Ng, and M. Peskin for useful discussions.  E.S. is grateful to the IAS for hospitality during parts of this project.  This research was supported in part by the National Science Foundation under grant PHY05-51164, by NSF grant PHY-0244728, and by the DOE under contract DE-AC03-76SF00515.


\appendix


\section{Appendix: Particle and String Production}\label{sec:appendix}

So far we have used the fact that a sector of sources $X$ may be produced through time dependent motion of the inflaton.  In this section, we will review and extend the computations of this effect.  We assume couplings of the inflaton $\phi$ to other degrees of freedom; this might be the generic situation and is necessary for reheating. For example, couplings of the form $\frac{1}{2}g^2\phi^2\chi^2$ endow the fields $\chi$ with a time dependent mass
\beq\label{masst}
m(t)^2=g^2\phi(t)^2+m_0^2 ~~
\eeq
as $\phi$ rolls.  In string theory, it is as common to have time dependent string tensions, for example of the form
\beq\label{tensiont}
T(t)^2=\eta^2M_P^2\phi^2(t)+T_0^2
\eeq
for some dimensionless coupling $\eta$ that depends on the string coupling $g_s$ and the shape and size ``moduli" of the extra dimensions.  Sufficiently close to $\phi=0$, we can approximate
\beq\label{v}
\phi(t)\approx vt
\eeq
with $v=\dot\phi$ the field velocity (of dimension 2).

A simple way to see how such effects arise in string theory is to consider the realization of scalar fields from the relative motion of branes (as well as various dual descriptions of this).
\begin{figure}[t!]
\begin{center}
\includegraphics[width=10cm]{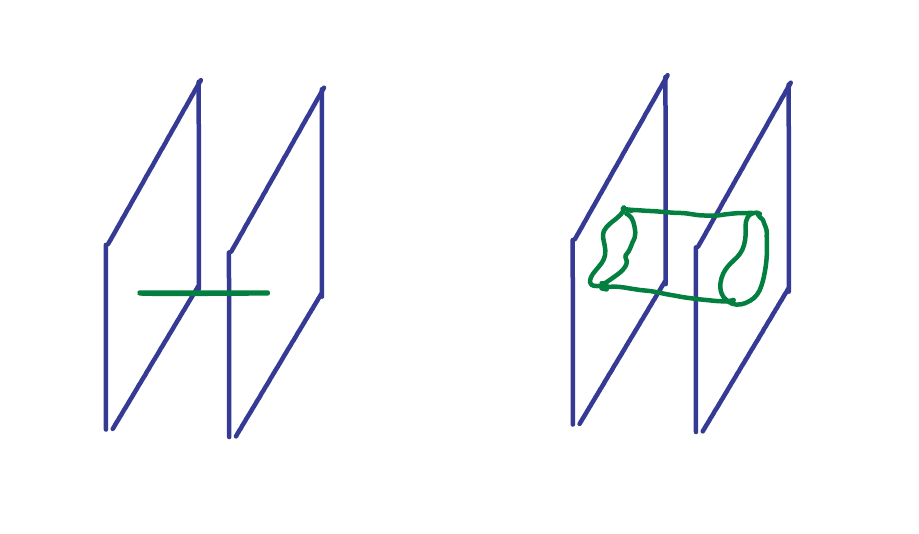}
\end{center}
\caption{Light degrees of freedom can be particles (left figure) or higher dimensional defects such as strings (right figure).}
\label{fig:branes}
\end{figure}
As two branes come together, strings stretching between them are particles in the worldvolume theory of the pair of branes which become light as in (\ref{masst})(\ref{v}).  Similarly, membranes stretching between two branes constitute strings in the worldvolume theory.  The tension $T(t)$ of these strings is given by the membrane tension times the distance between the branes, which is proportional to $\phi(t)$.  (In both cases, if the branes miss each other a nonzero mass $m_0$ or tension $T_0$ remains
at the minimal distance between them.)
In this section, we will first review the production of particles from (\ref{masst})(\ref{v}) and then generalize this to estimate the production rate of strings from (\ref{tensiont})(\ref{v}).

\subsection{More general mass functions}\label{subsec:massfunctions}

For some of our examples in the main text, we wish to consider a mass $m(\phi)$ for our produced source particles $\chi$ which becomes approximately constant (independent of $\phi$) at some point after production.  In this subsection, we describe a simple toy model with a few interactions depending on an additional heavy scalar $\phi_H$ which yields this behavior classically upon integrating out $\phi_H$.  This is closely related to a model described in \cite{Dong:2010in}.

The model has potential terms
\beq\label{heavyphi}
\left(M_H^2(\phi_H-\phi_0)^2+\phi_H^2(\phi-\phi_i)^2\right)\frac{\chi_i^2}{M_*^2}=m(\phi,\phi_H)^2\chi^2
\eeq
where the index $i$ refers to the production event which occurs at when the inflaton rolls through the point $\phi=\phi_i$ in field space.

Once the particles are produced, there is a number density $n_\chi$ (which, as we will see below, is of order $\dot m^{3/2}$ at the time of production).  This leads to energy density
\beq\label{rhochitwo}
\rho_\chi \sim m_\chi n_\chi \sim \langle \chi^2\rangle m(\phi,\phi_H)^2 .
\eeq
Equivalently,
\beq\label{chisquared}
\langle \chi^2\rangle\sim \frac{n_\chi}{m(\phi,\phi_H)}
\eeq

After the production event at $\phi=\phi_i$, at first $\phi-\phi_i$ is very small and the first term in (\ref{heavyphi}) freezes $\phi_H$ at $\phi_0$.  However, once $\phi-\phi_i$ becomes as large as $M_H$, the second term begins to dominate, and $\phi_H$ adjusts to a more energetically favorable value $\phi_H=\phi_{H*}[\phi_L])$ which depends on $\phi_L$.  Exactly as in the similar toy model of \cite{Dong:2010in}, the result of integrating out $\phi_H$ is that for $\phi-\phi_i\gg M_H$, the mass $m(\phi,\phi_{H*}[\phi_L])$ becomes constant.

This type of model -- and others more faithful to a concrete UV completion -- may provide a reasonable starting point for studies of the radiative stability of the scenario where $m(\phi)$ transitions to a constant.  Incorporating supersymmetry helps suppress loop corrections, though this alone does not prevent Hubble scale mass corrections.  Incorporating monodromy renders the corrections approximately periodic, restoring a discrete shift symmetry, but it is important to ensure that their amplitude is small enough to be viable.

\subsection{Bogoliubov Coefficients:  Particle case}
\label{subsec:particleprodI}

A particle with mass squared $m(t)^2=v^2t^2$ and frequency
$\omega(t)=\sqrt{v^2t^2+k^2}$ has a WKB wavefunction
\begin{equation}\label{wkbwavefunction}
\psi(t)\approx  {1\over{\sqrt{2\omega(t)}}}e^{-i\int^t\omega(t')dt'}
\end{equation}
which solves
\begin{equation}\label{parteom}
(-\partial_t^2 - v^2 t^2 -k^2 )\psi(t)=0
\end{equation}
to good approximation in the regime $\dot\omega\ll \omega^2$.

We start with this pure positive frequency wavefunction in the far past and evolve it to the far future, where it picks up a negative frequency term proportional to the Bogoliubov coefficient $\beta$ that determines the number of produced particles.  In terms of creation and annihilation operators, this means $a_{early}=\alpha a_{late}+\beta a^\dagger_{late}$.  So if we start in the vacuum, i.e. the state is the state killed by $a_{early}$, it is not killed by $a_{late}$ but instead is $e^{\beta (a^\dagger)^2/2\alpha}|late~vacuum>$ up to normalization.  (We are describing it in the Heisenberg picture, for which states do not evolve.) Taking the expectation value of $a_{late}^\dagger a_{late}$ reveals that $|\beta|^2$ is the number of produced particles.

In the above example, expanding
$\omega(t')\approx vt'+{k^2\over{2vt'}}$, the wavefunction at large $t$ becomes, doing
the integral in the exponent,
$$ \psi(t)\approx {1\over{\sqrt{2vt}}}e^{-i({t^2 v\over 2})}t^{-i{k^2\over {2v}}} $$
Continuing $t\to e^{-i\pi} t$, staying at large $|t|$ to preserve
the WKB approximation \cite{LL},  pulls out the Bogoliubov coefficient
$$\beta_k \sim e^{-\pi k^2/2v}. $$
Then the total number density of particles is $\int d^3k|\beta_k|^2\sim v^{3/2}$.
That is the dominant $k$ is $\sim\sqrt{v}$.

This was done in the case with the mass going through zero.  If instead $\omega^2=v^2t^2+\mu^2+k^2$,
we get
\begin{equation}\label{impact}
|\beta|^2\sim e^{-\pi(k^2+\mu^2)/v}
\end{equation}
So the number density of produced particles is model dependent, with $v^{3/2}$ being the upper bound.

\subsubsection{Associated scalar emission}\label{subsubsec:scalaremission}

In the bulk of this paper, we are interested in the gravitational wave emission from these produced sources.  An important consideration is the level of scalar emission that accompanies their production and interactions.  Here let us estimate this for the production event just reviewed.

A somewhat similar process was analyzed also in \cite{trapped}.  In the specific scenario worked out there, the scalar field dynamics is strongly affected by a very finely spaced set of points where different sectors of particles become light. In that case, because the scalar can lose energy to the many sectors of temporarily light fields, the Green's function for the scalar perturbation is not well approximated by its form in a free scalar field theory.  For our purposes in the present work, we may consider a somewhat simpler regime in which these events are spread out enough that the scalar Green's function is the standard one.  At the end of this section, we will verify that this is a self-consistent approximation.

Let us calculate the scalar radiation emitted from the particle production event, treated as a version of Bremsstrahlung.
This is similar to electromagnetic or gravitational Bremsstrahlung,
but with the source in this case obtained from the Born-Infeld action $-\int d\tau g\phi\sqrt{\dot t^2-{\dot{\vec x}}^2}$ to be
\beq\label{source}
\rho_\chi(x)=\sum_{n=1}^{N_\chi}\int d\tau\frac{1}{\gamma_n}\delta^{(4)}(x^\mu-x_n^\mu(\tau))
\eeq
where $\gamma_n=1/\sqrt{\dot t_n^2-\dot\vec x_n^2}=p_n^0/m_n$ where dot is derivative with respect to $\tau$.
From this, we obtain scalar radiation
\beq\label{scalarbrem}
\delta\phi_{rad}(x)=g\sum_{n=1}^{N_\chi}\int \frac{d^3\vec k}{(2\pi)^3}\frac{-m_n^2/p_n^0}{2|\vec k|}\frac{e^{-ik\cdot(x-x_n)}}{k\cdot p_n}+c.c.
\eeq
where we have summed over multiple $\chi$ pairs produced in the event.

We can simplify this by taking into account the fact that the particle slows down quickly; $p^0\approx m$.  Putting that in and considering a production event where $N_\chi$ particles are all produced at $t=0$ gives
\beq\label{scalarpert}
\delta\phi(x)\sim g \int_H^{\sqrt{\dot\phi}} \frac{d^3\vec k}{\vec k^2}e^{-ik\cdot x}\sum_{n=1}^{N_\chi}e^{i\vec k\cdot{\vec x}_n}\ .
\eeq
We have put the lower limit at $H$ because below that we cannot use the flat spacetime analysis, and the upper limit at $\sqrt{\dot\phi}$ because the particle production process takes a time of order $1/\sqrt{\dot\phi}$~~\footnote{In sec.~\ref{subsubsec:GWprod} we consider the case in which $\chi$ particles are coupled to $\phi$ for a time shorter than an Hubble time, which, as we will see, leads to a suppression of the amount of $\delta\phi$ radiation produced at low frequencies.}.  Note that here $\vec k$ is the physical momentum, which we could call $\tilde{\vec k}/a(t)$ in terms of the comoving momentum $\tilde{\vec k}$.

Let us determine the expectation value of $\delta\phi(x)^2$:
\be\label{eq:deltaphi1}
\langle\delta\phi(x)^2\rangle\simeq g^2 \int_H^{\sqrt{\dot\phi}} \frac{d^3k}{(2\pi)^3}\frac{1}{k^2}\int_H^{\sqrt{\dot\phi}} \frac{d^3k'}{(2\pi)^3}\frac{1}{k'{}^2}e^{-i(k-k')\cdot x}\langle\sum_{n=1}^{N_\chi}e^{i\vec k\cdot{\vec x}_n}\sum_{n'=1}^{N_\chi}e^{-i\vec k\cdot{\vec x}_{n'}}\rangle\ .
\ee
Let us evaluate the term on the right inside the expectation value. Since particles are uncorrelated, this is proportional to $\delta_{n,n'}$. This leads to
\be
\langle\sum_{n=1}^{N_\chi}e^{i\vec k\cdot{\vec x}_n}\sum_{n'=1}^{N_\chi}e^{-i\vec k\cdot{\vec x}_{n'}}\rangle\sim\sum_{n=1}^{N_\chi}\langle e^{i(\vec k-\vec k')\cdot{\vec x}_n}\rangle\sim N_\chi H^3\int d^3x\; e^{i(\vec k-\vec k')\cdot{\vec x}_n}\sim N_\chi H^3(2\pi)^3\delta^{(3)}(\vec k-\vec k')\ ,
\ee
where in the next to last step we have approximated the summation with the continuum limit and taken into account of the fact that the integral is limited to an order one Hubble patch. Plugging back into~(\ref{eq:deltaphi1}) and carrying one of the two $k$ integrals through the delta function, we have
\be\label{eq:deltaphi2}
\langle\delta\phi(x)^2\rangle\simeq g^2H^3 N_\chi \int_H^{\sqrt{\dot\phi}} \frac{d^3k}{(2\pi)^3}\frac{1}{k^4} \ .
\ee
We are interested in the resulting curvature perturbation at horizon crossing, where $\zeta\sim H\delta\phi/\dot\phi$.
Since fluctuations in $\delta\phi$ redshift as $H/\omega$, with $\omega$ being their frequency of emission, for each logarithmic frequency bin we obtain
\be
\frac{d P_\zeta}{d\log \omega} \sim \frac{H^4}{\dot\phi^2} g^2N_\chi\left(\frac{H}{\omega}\right)^3\ .
\ee
Because of the redshift, the contribution to $\zeta$ is dominated by $\delta\phi$ fluctuations emitted directly at $\omega\sim H$, leading to
\be\label{eq:scalar_constrint}
 P_\zeta \sim \frac{H^4}{\dot\phi^2} g^2N_\chi\lesssim 10^{-10} \ ,
\ee
where the last relation follows from the normalization.  For large $g^2 N_\chi$, we need a sufficiently small inflationary scale in order to match the normalization.  This is an additional constraint, but one that can be satisfied for the examples given above.

It is useful to obtain the same result as above in a way that is more similar to the way we obtain our result for gravitational waves. The energy density is $\delta\phi$ waves due to Bremmstrahlung is given by
\be
\frac{d\rho_\phi}{d\log\omega}\sim \omega^2\frac{d\langle\delta\phi^2\rangle}{d\log \omega} \sim n_\chi \omega \frac{d E}{d\omega}\ ,\quad\Rightarrow\quad\frac{d\langle\delta\phi^2\rangle}{d\log \omega}\sim \frac{n_\chi}{ \omega} \frac{d E}{d\omega}\ ,
\ee
$dE/d\omega$ follows the same expression as for gravitational waves with $(E/\mpl)^2$ replaced by $g$:
\be
\frac{d E}{d\omega}\sim g^2\ ,
\ee
leading to
\be
\frac{d\langle\delta\phi^2\rangle}{d\log \omega}\sim \frac{g^2 n_\chi}{\omega}  \sim \frac{g^2 H^3 N_\chi}{\omega} \ ,
\ee
which nicely agrees with what found above in (\ref{eq:deltaphi2}). We can turn this into an estimate for $\zeta^2$,
\beq\label{scalarpmat}
\zeta^2 \sim  \frac{g^2 n_\chi}{H}  {H^2 \over \dot \phi^2} \sim  \frac{g^2 f }{\epsilon} \frac{H}{E}\ ,
\eeq
and notice that this also implies
\be\label{eq:NphiNchi}
N_\phi\sim g^2\ N_\chi\ ,
\ee
with $N_{\phi,\chi}$ representing the number of respectively $\phi$ and $\chi$ particles in an Hubble patch.
Finally, let us come back to the check we mentioned above, that we may use the free scalar Green's function in
(\ref{scalarbrem}) in a useful range of parameters.  To study this, we should compare the scalar perturbations $\delta\phi$ with the distance $\Delta\phi$ between particle production events. If the ratio $\delta\phi/\Delta\phi$ is less than 1, then the perturbations do not typically lose energy to $\chi$ production. By taking $\delta\phi\sim\langle\delta\phi(x)^2\rangle$ and using (\ref{eq:deltaphi2}), this ratio is
\beq\label{phipertcomp}
\frac{\delta\phi}{\Delta\phi}\sim \frac{H N_{hits}}{\dot\phi}\delta\phi \sim g \frac{N_\chi^{1/2} H^2}{\dot\phi}\cdot N_{hits}\lesssim   P_\zeta^{1/2}N_{hits}\ ,
\eeq
where $N_{hits}$ is the number of particle production events per Hubble time and where in the last passage we have used the constraint from (\ref{eq:scalar_constrint})~\footnote{One could use the upper bound of the integral in (\ref{eq:deltaphi2}) to estimate $\delta\phi/\Delta\phi$ to ensure that even the most high-energy modes are not affected by particle production.  This would lead to multiply the last term in (\ref{phipertcomp}) by a factor of $(H^2/\dot\phi)^{1/4}\ll 1$, which leads to an even milder constraint on $N_{hits}$.}.  This ratio can be small consistently with a large $N_{hits}$, consistent with approximate scale invariance.  (Because the events are discrete, there will be structure on small scales in the power spectrum, but for sufficiently finely spaced events these oscillations wash out.)

\subsection{String Case}

It is interesting to generalize this to strings with a time dependent tension $T(t)$ \cite{stringprod}.\footnote{We thank J. Polchinski for very useful discussions.}  In general this is not simply a sum over string oscillator states treated as particles \cite{Gubser:2003vk}, since the tension can vary rapidly enough that the string cannot causally adjust to maintain its oscillator configuration.  We find that large pairs of loops are formed, but  in some cases the relative oscillations between the strings and anti strings making up the pair are displaced from each other by a short distance, roughly of order $\dot T^{-1/3}$.  (This is required by causality.)  They generically oscillate relative to each other, and may quickly decay into smaller loops of that size, though joining transitions forming longer loops also occurs at some level, depending on the density \cite{Mitchell:1987th}.   There may also be straighter pairs of loops formed more quickly, with a different decay pattern.  We leave that case for future work.

\subsubsection{Circular Loops}
\label{subsubsec:cirloops}

Let us start by considering the simplest configuration (circular loops).
The generalization of (\ref{parteom}) for a
circularly symmetric loop's wavefunction $\Psi(r,t)$ is
\begin{equation}\label{stringeom}
(-\partial_t^2 + \partial_r^2 - b^2 t^2 r^2-k^2 )\Psi(r,t)=0
\end{equation}
where we took the time-dependent tension $T(t)$ to behave as $T(t)^2=b^2 t^2$; i.e. $b=\dot T$.

This can be derived from the Hamiltonian constraint in the string worldsheet theory
\begin{equation}\label{wsaction}
S_{string}=\int d\sigma d\tau \sqrt{-g}T(t)
(g^{\alpha\beta}\partial_\alpha X^M\partial_\beta X^NG_{MN})
\end{equation}
given in terms of the spacetime metric
\begin{equation}
G_{MN}dx^Mdx^N=-dt^2+dr^2+r^2d\theta^2+d\vec x_\perp^2
\end{equation}
Varying (\ref{wsaction}) with respect to the metric produces the constraints
\begin{equation}\label{constraints}
0\equiv T_{\alpha\beta}=-T(t)(\partial_\alpha X^M\partial_\beta X^NG_{MN}-\frac{1}{2}g_{\alpha\beta}\partial_\gamma X^M\partial^\gamma X^N G_{MN})
\end{equation}
We are interested in a simple circularly symmetric configuration, with $\theta=\sigma$ and $r, \vec x_\perp$ and $t$ being functions of $\tau$.  For simplicity we are ignoring motion of the string in the $r,\theta$ plane but this could be included.
The $T_{\sigma\tau}=0$ constraint is solved by $g_{\sigma\tau}=0$.  The others are solved by
\begin{equation}\label{solconstraints}
-\dot t^2+\dot r^2+\dot{\vec{x}}_\perp^2+r^2 = 0
\end{equation}
Now writing this in terms of momenta, using $p_t=\frac{\partial {\cal L}}{\partial{\dot t}}=-T(t)\dot t
=i\partial_t$ (the last being the representation of the momentum operator in position space), and similarly for the other coordinates and momenta, yields
(\ref{stringeom}).

In our problem (\ref{stringeom}), $br$ is like $v$ in (\ref{parteom})
except there is also the $\partial_r^2\Psi$ term.
However, if we work at $r\gg t$ we have a regime where this term is subdominant in (\ref{stringeom}), basically
because the derivatives with respect to $r$ pull down factors of $1/r$, which is smaller than
$1/t$.  That is, taking (\ref{wkbwavefunction}) with now $\omega^2(t)=b^2 t^2 r^2 + k^2$, we get
\beq
 \ddot\Psi\approx-\omega^2(t)\Psi=-(b^2r^2t^2+k^2)\Psi
\eeq
but
\beq
\Psi'' \approx -(b^2 t^4)\Psi
\eeq
For $r\gg t$, this is subdominant to the first, $b^2r^2t^2$ term in (3).  It can also
be subdominant to the $k^2$ term in (3) for the dominant $k\sim \sqrt{br}$.  In order
for the large t expansion to be valid, we needed $\dot\omega/\omega^2\sim 1/(brt^2)\ll 1$.
This seems to be consistent with the above analysis as long as $r$ is large enough, $r\gg b^{-1/3}$.

According to this calculation, the time-varying tension $T=bt$ can produce many pairs of large loops, because we get from this
\beq\label{betacoeff}
|\beta|^2\sim e^{-\pi k^2/br}
\eeq
This was all done in the case that the tension goes through zero.  One can also study similarly the case where it does not, giving (c.f. (\ref{impact}) with $\mu=T_{min}r$)
\begin{equation}
|\beta|^2\sim e^{-\frac{\pi}{b}(\frac{k^2}{r}+T_{min}^2r)}
\end{equation}

Although the loops can be large, the distance between them is much smaller, consistent with causality.  Moreover, they move very slowly overall:  the peak momentum from (\ref{betacoeff}) is $k_*\sim M v\sim \sqrt{br}$.   Using that $M\sim b^{2/3}r$ at the time $t\sim b^{-1/3}$ of production, we find that the strings have a relative velocity
\beq\label{vstring}
\Delta v_{string}\sim \frac{1}{b^{1/6}r^{1/2}}
\eeq
Therefore the time it takes to separate the two members of the pair by a distance $\dot T^{-1/3}=b^{-1/3}$ is
\beq\label{tsep}
t_{separation}\sim b^{-1/6}r^{1/2}
\eeq
This is much greater than the timescale $\dot T^{-1/3}\sim b^{-1/3}$.

This is important because the latter is the timescale for relative oscillations of the string in more general configurations.
The circular configuration is not generic, and it is interesting to consider the problem more generally by studying the string path integral \cite{stringprod}.  This yields a similar result, but  includes cases in which the two members of the pair of loops oscillate relative to each other at a relative distance $b^{-1/3}\sim \dot T^{-1/3}$.  The simplest string-theoretic examples which develop light strings (which are sometimes called ``tensionless string theories") do not have a small coupling suppressing interconnection of the strings or decay into scalar modes.  As a result, the pair production of large strings in these cases is quickly followed by the pair breaking up into a ring of smaller loops.  This is the situation analyzed in \S\ref{subsec:stringring}.   Another case closer to the circular string example appears to be where strings are produced without relative oscillations, in a non-adiabaticity timescale of order $\dot M/M^2\sim 1/\sqrt{r\dot T}$ where $M\sim r T$ is the mass of the long string.  We leave the analysis of this case for future work.

\begingroup\raggedright\endgroup

\end{document}